\documentclass[12pt]{article}
\usepackage{amsfonts,amssymb,amsmath,amscd}
\usepackage{epsf}

\def\d{\partial}

\def\d{\partial}
\def\inv{^{-1}}

\def\({\left(}
\def\){\right)}
\def\[{\left[}
\def\]{\right]}

\def\({\left(}
\def\){\right)}

\def\d{\partial}
\def\inv{^{-1}}

\def\beq{\begin{equation}}
\def\eeq{\end{equation}}
\def\bea{\begin{eqnarray}}
\def\eea{\end{eqnarray}}
\def\bq{\begin{quote}}
\def\eq{\end{quote}}

\def\d{\partial}

\def\({\left(}
\def\){\right)}
\def\e{e^}
\def\tr{{\rm{tr}}}     
\def\g5{\gamma_5}
\def\gappeq{\mathrel{\rlap {\raise.5ex\hbox{$>$}}
{\lower.5ex\hbox{$\sim$}}}}

\def\lappeq{\mathrel{\rlap{\raise.5ex\hbox{$<$}}
{\lower.5ex\hbox{$\sim$}}}}

\def\Toprel#1\over#2{\mathrel{\mathop{#2}\limits^{#1}}}

\begin{document}

\pagestyle{empty}
\begin{flushright}
{CPHT-RR 048.0804}\\ 
{\ttfamily hep-th/0410180}\\ 
\end{flushright}
\vspace*{5mm}
\begin{center}
{ \Huge Tachyon Potential in a Magnetic Field 
 \\
 \vspace*{3mm} \Huge with Anomalous Dimensions}\\
\vspace*{19mm}
{\large Pascal Grange} \\
\vspace{0.5cm}

 {\it Centre de physique th{\'e}orique de l'{\'E}cole polytechnique,\\
 \vspace*{0.2cm}route de Saclay, 91128 Palaiseau Cedex, France}\\
\vspace{0.5cm} 
{\tt{ pascal.grange@cpht.polytechnique.fr}}\\
\vspace*{3cm}

\vspace*{1.5cm}  
{\bf ABSTRACT} \\ \end{center}
\vspace*{5mm}
\noindent
 
 Products defined in the context of noncommutative gauge theory allow for an interpolation
 between exact results on tachyon potentials at zero and large
background $B$-fields. Techniques for computations of effective
actions are transposed from the framework of gauge theory to the
framework of boundary string field theory, resulting in deformations
of the noncommutative tachyon potential by anomalous dimensions.

\vspace*{0.5cm}
\noindent

\begin{flushleft} 
October 2004 
\end{flushleft}

\vfill\eject

\setcounter{page}{1}
\pagestyle{plain}

\section{Introduction}

Open strings propagating in a definite closed-string background give rise to a
path integral whose classical action consists of two terms. The first
one is an integral over the whole world-sheet $\Sigma$ encoding the
closed-string background, while the second one is an integral over the
boundary encoding the coupling of open strings: 
$$Z=\int DX \exp{\Big{(}} -S_\Sigma[X]-S_{\d\Sigma}[X] {\Big{)}}.$$
When $S_\Sigma$ is taken to be the coupling of $\Sigma$ to the flat
space-time metric, as well as to a constant $B$-field, and $S_{\d\Sigma}$ to be the integral of the gauge
field carried by open strings, namely 
$$S_{\Sigma}[X]=\int_\Sigma \(\delta_{\mu\nu}dX^{\mu}\wedge\ast dX^{\nu} +B\),$$
$$S_{\d\Sigma}[X]=\int_{\d\Sigma} A,$$
the saddle-point approximation to the path integral above allows for a
stringy derivation of the Born--Infeld action~\cite{FT1,FT2}. We will dwell
on the case of fully Neumann boundary conditions and Abelian gauge
field (a single space-filling D-brane is present).\\

Going beyond this approximation amounts to taking derivatives of the
field strength into account. Andreev and Tseytlin~\cite{corrections} worked out the
first derivative corrections, and Wyllard brought more systematics~\cite{Wyllard}
into the computation of those corrections at higher orders, both in
the Born--Infeld and in the Wess--Zumino couplings. The Seiberg--Witten limit~\cite{SW} of
string theory,  where noncommutative field theory becomes a valid
description of the gauge theory along the brane, subsequently provided
lots of predictions for corrections from the very structure of
star-products~\cite{DMS}. Those predictions, encoded~\cite{DT,Liu} in $n$-ary products $\ast_n$ were proven to be consistent with the
work of Wyllard at low orders in derivatives, and then derived from
commutative string theory~\cite{Mukhi,mezigue}.\\

These encouraging results obtained in the Seiberg--Witten limit invited to
go beyond this limit, investigating the contribution of anomalous
dimensions. This led to the definition of deformed star-products $\tilde{\ast}_n$ from
the computation of path integrals, that in turn induced deformations
of noncommutative gauge theory~\cite{MSbeyond,Gbeyond}.\\

Another physical sector carried by open strings
comes from the tachyon field coupled to the boundary, 
$$S_{\d\Sigma}[X]=\int_{\partial\Sigma}T\Big{(}X|_{\partial\Sigma}(\tau)\Big{)}d\tau.$$
 This is a
different physical framework from the one of the gauge sector, with
important non-perturbative outputs regarding the vacuum of string
theory through the tachyon condensation picture~\cite{condensation}. Boundary string field theory~\cite{Shatashvili1,Shatashvili2,GS} gave rise to successful checks of tachyon condensation~\cite{KMM1,KMM2}. Investigations of noncommutative tachyons were performed in the {\hbox{large-$B$}} limit~\cite{DMR,WittenNT}. From
the technical point of view, similarities may be noted with the
results on the gauge sector, as far as noncommutative field theory is
concerned. Star-products have indeed been proven to be relevant for
 the expression of the tachyon effective actions for the bosonic
 string in works by Cornalba
 and Okuyama~\cite{Cornalba,Okuyama}. The model was in fact exactly
 solved for a quadratic tachyon field, and shown to be consistent with
 tachyon condensation.\\

However, the step of the deformation of star-products still has to be
completed, in order to really parallel the insights we got on the
gauge sector from noncommutative field theory. This is the purpose of
the present note. I shall first perform computations with a general
tachyon field, thus recognizing the pattern of Abelian gauge
theory. Differences coming from the absence of fermions and induced form gradings
 will be outlined. The result will be analogous to the previously
 known noncommutative tachyon effective action, except for the
 deformation of star-products, and the restoration of the kinetic
 term. As a check, I shall recover the exact solution for a quadratic
 tachyon potential, just by bounding the order of allowed derivative
 corrections.\\

\section{Deforming the partition function}

Let $\Sigma$ be the world-sheet with the topology of a disk.
The boundary is coupled to  the tachyon field $T$.
Expanding the factor containing the boundary interaction in powers
of $T$ in the partition function
$$Z(T)=\int DX e ^{-S_{\Sigma}[X]-\int_{\d\Sigma}T(X(\tau))d\tau}$$
 naturally leads to integrals over various ways of
inserting any number of tachyons along the boundary of the disk. In
fact we are computing a generating function for such amplitudes,
corresponding to smearing tachyons over the boundary.\\

The computation of the path integral $Z(T)$ is formally similar to the
one performed to obtain star-products from ordinary string theory. Of
course  we do not have fermionic zero modes in the present context to
build the grading of differential forms coupled to Ramond--Ramond form fields, but the expansion of the spatially-varying tachyon field
makes fluctuations of scalars appear at various powers, whose
contractions with the scalar propagator provide the result of the
path integral. Namely, let us split the scalar field into zero mode and
fluctuating part:
$$X^\mu(\sigma)=x^\mu+\xi^\mu(\sigma),$$
and write subsequently
 $$T(X(\sigma))=T(x)+\sum_{p\geq 1}\frac{1}{p!}\xi ^{\mu_1}(\sigma)\dots \xi
^{\mu_k}(\sigma)\partial_{\mu_1}\dots\partial_{\mu_k} T(x).$$ 
The integrand splits into a Born--Infeld contribution and 
an expectation value of the boundary insertion.
 $$Z(T)= \int dx\,\sqrt{\det(1+B)}\int D\xi e ^{-S_{\Sigma}-S_{\partial\Sigma}}. $$  

 Expanding  the integrand to a given
order in the tachyon field produces monomials in the fluctuations $\xi^\mu$ of
the coordinates. Those fluctuations that contribute to the path-integral 
allow for Wick contractions that do not connect scalars located at the
same $\sigma$. The simplest of such contributions comes from the
quadratic order in the tachyon and reads:
 $$\int d\sigma_1\int d\sigma_2 \sum_{k\geq 1} \(\frac{1}{k!}\)^2 \xi ^{\mu_1}(\sigma)\dots \xi^{\mu_k}(\sigma_1)\xi ^{\nu_1}(\sigma_2)\dots \xi^{\nu_k}(\sigma_2)\partial_{\mu_1}\dots\partial_{\mu_k} T(x)\partial_{\nu_1}\dots\partial_{\nu_k} T(x).$$
One of the two integrations is dealt with using translation invariance, one
of the $1/k!$ factors is compensated by the number of possible
contractions with the two-point function of scalar fields on the disk (regularized by a small parameter $\epsilon$):
$$D^{\mu\nu}(\sigma)=~\alpha '\(\frac{\theta
    ^{\mu\nu}}{2\pi\alpha'}\log\(\frac{1-e ^{-\epsilon+i\sigma}}{1-e
      ^{-\epsilon-i\sigma}}\)+G^{\mu\nu}\log|1-e ^{-\epsilon+i\sigma}|^2\).$$
This propagator was used in~\cite{MSbeyond,Gbeyond} in order to dress noncommutative field theory with the contribution of anomalous dimensions. The simpler noncommutative field theory obtained in the Seiberg--Witten limit corresponds to keeping only the first term. As far as combinatorial problems are concerned, the counting of contractions is performed in the same way, whether or not the second  term is included. It is explained in~\cite{mezigue} how to recursively obtain $\ast_k$-products for the contribution of order $k$ in the field strength. These products therefore found a derivation from commutative string theory after their first appearance from geometric consistency conditions in noncommutative gauge theory~\cite{DT,Liu}. It was noticed that the recursion worked for arbitrarily high order $k$, and was only stopped by dimensionality (antisymmetry of the finite algebra of fermionic zero modes). In the present context of tachyon fields, the contribution up to quadratic order to the path integral reads 
$$Z(T)=\int d x\,\sqrt{\det(1+B)}\(1+T+\frac{1}{2!}T\tilde{\ast}_2 T+\dots\),$$
where $\tilde{\ast}_2$ denotes the deformation of the binary differential operator $\ast_2$:
$$ \tilde{\ast}_2=\frac{\Gamma(1+2\d G\d')}{\Gamma(1-\d\theta\d'+\d G\d')\Gamma(1+\d\theta\d'+\d G\d')},\;\;\;\;\;\;\;\;\;\;\;\;{\ast}_2=\frac{\sin(\d\theta\d'/2)}{\d\theta\d'/2}.$$  
The first term in $Z$ comes from the zero-tachyon contribution, and the
first order in $T$ does not receive corrections because those could only come
from self-contractions of scalars.  
The recursive computations of~\cite{mezigue,Gbeyond} can be reproduced
in the same fashion as above, leading to:
$$Z(T)= \int dx\,\sqrt{\det(1+B)}\(1+T(x)+\sum_{k\geq 2}
\frac{1}{k!}\tilde{\ast}_k[T^k(x)]\),$$
where the higher-degree deformed $k$-ary operators contribute:
$$\tilde{\ast}_k=\int_0^1 d\tau_1\dots\int_0^1 d\tau_k \prod_{1\leq i<j\leq k}\exp\left\{i\pi a_{ij}(2\tau_{ij}-\epsilon(\tau_{ij}))\right\}(2\sin(\pi\tau_{ij}))^{2t_{ij}}.$$

We symbolically denote the result by 
$$Z(T)=\int d x\,\sqrt{\det(1+B)}\exp(-\tilde{\ast}T(x)).$$
Taking the anomalous dimensions of the tachyons to zero reduces the
deformed $k$-ary star-products to the ordinary ones $\ast_k$
~\cite{Liu}. Taking $B$ to be large then yields the expression derived
in~\cite{Cornalba}, as it should.

\section{Consistency check for a quadratic tachyon field}

Let us consider the exactly solvable model with  a quadratic tachyon field
$$T(X):= a + u_{\mu\nu} X^{\mu}X^{\nu},$$
studied by Witten without a $B$-field~\cite{Witten}, where $u_{\mu\nu}$ is a symmetric tensor. 
The partition function in the closed-string background we considered above was obtained by Okuyama~\cite{Okuyama},
who computed the propagator and restricted it to the boundary. Such a computation is analogous to 
 the derivation of the  Born--Infeld action in~\cite{FT2}, but the saddle-point approximation 
is in fact exact. This result should be reproduced by our general formula. In the present case, 
the derivative expansions are cut at quadratic order, making some solidarity appear between the order in the tachyon field and the order in derivatives.
 Fortunately, Wyllard studied~\cite{Wyllard} the relevant analogous terms in boundary-state computations for Wess--Zumino couplings. We will see that our check reduces to a translation work. We will namely have to provide a dictionary 
from the language of $2n$-derivative corrections to $2n$-forms, to the language of $2n$-derivative corrections to $2n$-th powers of the scalar field $T$.\\

 The work~\cite{Wyllard} addressed the problem of the
 $2n$-form $2n$-derivative terms in the {\hbox{Wess--Zumino}} couplings
 to a space-filling D9-brane in the background of a flat metric and
 constant $B$-field. It is of the same class as the problem at hand,
  since the interactions in the path integral also consist of a
 boundary coupling:
$$\int d\tau d\theta D\phi^\mu A_\mu(\phi)=-\int d\tau d\theta\sum_{k\geq 0}\frac{1}{(k+1)!}\frac{k+1}{k+2}D\tilde{\phi}^\nu \tilde{\phi}^{\mu}\tilde{\phi}^{\mu_1}\dots\tilde{\phi}^{\mu_{k}}\d_{\mu_1}\dots\d_{\mu_{k}}F_{\mu\nu}(x)$$
$$-\int d\tau(\tilde{\psi}^\mu\psi_0^\nu+{\psi}_0^\mu\psi_0^\nu)\sum_{k\geq 0}\frac{1}{k!}\tilde{X}^{\mu_1}\dots\tilde{X}^{\mu_k}\d_{\mu_1}\dots\d_{\mu_k}F_{\mu\nu}(x).$$
As discussed above, the only contractions in the
 computations of~\cite{Wyllard} that contribute in our bosonic framework are
 those between scalars. We are going to  map the computation of the
 Wess--Zumino couplings to our context,
$$\int d\tau d\theta D\phi^\mu A_\mu(\phi)\mapsto\int T(X(\tau)) d\tau,$$
 by replacing the field strengths $F_{\mu\nu}$ in those couplings  with tachyon fields $T$, and cancelling any term that does not carry two
 zero-modes for each field strength (since those involve
 fermions). Finally, we have to replace any product of fermionic zero-modes by 1,
  thus destroying the graded structure and allowing for terms of
 arbitrarily high degree in $T$.\\

The simplification that occurs,  due to the quadratic nature of the
 tachyon field, comes from the low number of derivatives that can act
 on it. A correction of order $n$ in $u_{\mu\nu}$ namely has to come from the
 expansion at order $n$ in the tachyon field, with two derivatives
 acting on each of the factors, as:
$$\sum_{p\geq 2}\int\(\prod_{i=1}^p d\sigma_i\)\frac{1}{(2!)^p}\(\prod_{i=1}^p
\xi ^{\mu_1}(\sigma_i)\xi ^{\mu_2}(\sigma_i)\d_{\mu_1}\d_{\mu_2}T(x)\).$$
These corrections belong to the formal class of $2n$-form
$2n$-derivative corrections derived in~\cite{Wyllard} with the
 analogies described above: the two-form $F$ is replaced with $T$, and the
term $\d_{\rho_1}\d_{\rho_2}F_{\mu_1\mu_2}$ therefore reduces to $u_{\rho_1\rho_2}$.
  With these rules, the analogous object to the matrix-valued two-form called ${\rm{S}}$ in
section 3 of~\cite{Wyllard} (taking the form of the curvature of non-symmetric gravity) is just the tensor $u_{\mu\nu}$ (since the graded structure of the algebra of forms disappeared):
$$ {\rm{S}}=\d_{\rho_1}\d_{\rho_2} F +2 \iota_{h}\d_{\rho_1} F\wedge \d_{\rho_2}
F\longrightarrow u_{\rho_1\rho_2},$$
where $\iota$ is the inner multiplication by vector fields and $h$ is
the inverse of $g+B$, with symmetic and antisymmetric parts denoted as
 $G$  and $\theta$:
$$ h^{\mu\nu}=: G^{\mu\nu}+\theta^{\mu\nu}.$$ 
Terms of the form
$$h^{\nu_1\nu_2}\d_{\rho_1}F_{[\mu_1|\nu_1} \d_{\rho_2}
F_{\mu_2]\nu_2}$$
disappear, because they come from integration of fermionic
fluctuations.
At this point we are instructed to perform contractions that would
yield the deformed $\tilde\ast_n$ differential operators if $T$ was not annihilated
by higher-order derivatives. What we are going to obtain is a
truncation of these operators, in which each of the $n$ factors is
acted on at most twice. At fixed order $n$ in $u$, it is exactly what
is called $W_{2n}$ in~\cite{Wyllard}
$$S_{WZ}=\int C\wedge e^F\wedge\(1+\sum_{n\geq 2} W_{2n}\) ,$$
 with all orders $n$ allowed (they are
only formally defined for $n>5$ in the gauge-theory context because of the
dimension of the world-volume and the grading
induced by fermionic zero modes). With these notations the first few terms
are:
$$W_4= \frac{\zeta(2)}{2}\tr[(hu)^2],$$
$$ W_6=    \frac{\zeta(3)}{3}\tr[(hu)^3],$$
$$ W_8= \frac{\zeta(4)}{4}\tr[(hu)^4]+\frac{1}{2}\(\frac{\zeta(2)}{2}\)^2\(\tr[(hu)^2]
\)^2,$$
so that the contribution of all the corrections yields an
exponential factor of
$$ \exp\( \sum_{n\geq 2}\frac{\zeta(n)}{n} \tr[(hu)^n]\).$$
Since $u_{\mu\nu}$ is symmetric, the traces can be expressed either as
contractions with $h$ or $h^t$, so that we can make contact with the
more symmetric notations of~\cite{Okuyama}: 
$$\tr[((G \inv+\theta)u)^n]=\tr[(hu)^n]=\tr[(u ^t h^t)^n]= \tr[(u
(G\inv-\theta))^n].$$
We are ready to use the $\Gamma$-function identity
$$-e ^{-\gamma z}\Gamma(1-z)=\exp\(\sum_{n\geq 2}\frac{\zeta(n)}{n}
z^n \),$$
which was actually written as a formal remark by Wyllard, even if large-$n$ contributions were cut-off in the gauge sector. The insertion of tachyons on the boundary gives rise to the right  physical arena for the application ..of this formula, since all the terms of any given order $n$ do contribute.  
According to the above identity we obtain
$$Z=e ^{-a}\exp\(\frac{1}{2}\sum_{n\geq 2}\tr[((G
\inv+\theta)u)^n]\)\exp\(\frac{1}{2}\sum_{n\geq 2}\tr[((G
\inv-\theta)u)^n]\)$$
$$=\e{-a}e ^{-\gamma\tr{G\inv u }}\sqrt{\det\Gamma(1-(G
  \inv+\theta)u)}\sqrt{\Gamma(1-(G \inv-\theta)u)},$$
in agreement with the Gaussian determinants evaluated in~\cite{Okuyama}. Our proposal has therefore passed a first test. Let us also note that the formal limit $\theta\rightarrow 0$ of $\tilde{\ast}_2$ (containing only contributions from anomalous dimensions and no from noncommutativity), can be recognized in the partition function derived in~\cite{CFGNO}, where the $\beta$-function method~\cite{KS} is carried on up to cubic order.

\section{Conclusions}

Derivative expansions of open-string effective actions may be obtained from algebraic rules 
applied to fluctuations of world-sheet coordinates. These rules embody the presence of D-branes
 as boundary states, and are dictated by the closed-string background. When the latter contains 
a large constant $B$-field, noncommutativity introduces some more structure into the form of the effective action, 
under the disguise of star-products.

 These products are then subject to deformations coming from short-distance singularities in the operator product expansions. We have seen that this program can be completed in the case of tachyon insertions on the boundary on the world-sheet, as well as in the case of photon insertions, where it was originally worked out. Moreover, objects such as $k$-ary products $\ast_k$ or $\tilde{\ast}_k$ with $k\geq 5$, which were defined only formally in the context gauge theory, contribute to the partition function computed above. 
Even though the Seiberg--Witten map is intimately tied to gauge invariance and gave rise to the definition of the undeformed products~\cite{DT,Liu}, all these products and their deformations find a home in the effective theory of tachyons. This makes them more relevant for open-string effective actions in the background of a $B$-field.\\

We obtained a closed expression for the partition function, which dresses the former results with anomalous dimensions 
only through deformations of the star-products. Moreover, our expression gives back the exact solution in the case of a quadratic tachyon field.
 As for the effective action, taking anomalous dimensions into account will always generate a kinetic term. 
The effective action is indeed related to the partition function through~\cite{GS}
$$S(T)=\(1-\int d^2x\,\beta(T(x))\frac{\delta}{\delta T(x)}\) Z(T),$$
so that, in the case of quadratic tachyon field, the effective action is  obtained by acting on the partition function with derivatives on the space of parameters $a$ and $u_{\mu\nu}$, as ~\cite{Okuyama,Witten}
$$S(a,u)=\(1+{\rm{tr}}(G^{-1}u)-a\frac{\d}{\d a}-{\rm{tr}}\(u\frac{\d}{\d u}\)\)Z(a,u).$$
where the second term was neglected in~\cite{Cornalba} because the $\beta$-function of the tachyon field $T$ is simply linear in $T$ whenever anomalous dimensions are neglected. Once restored, it will generate a kinetic term in the effective action. As for the potential, we are left with the potential predicted by Gerasimov and Shatashvili, again dressed with modifield star-products:
$$V(T)=(1+T)\tilde{\ast} e^{-\tilde{\ast}T}.$$     
The zero-momentum limit of the deformed star-products cancels both the contribution of anomalous dimensions
and those of Moyal phases. We therefore notice, following the remark of Dasgupta, Mukhi and Rajesh~\cite{DMR}, that universality of the tachyon potential~\cite{universality} is not violated by the deformations we exhibited.
Implications for noncommutative solitons~\cite{GMS} could come from field redefinitions mapping the kinetic term to the ordinary one. It seems difficult to study solitons with anomalous dimensions included in the noncommutative picture, but on the other hand, recent studies of the toy-model of noncommutative $p$-adic string~\cite{GhoshalSolitons,padic,towards}could provide a hint. Allowing the coordinates on the boundary of the world-sheet to take values in the field ${\rm{\bf Q}}_p$
 of $p$-adic numbers led to an exactly solvable field theory for open string tachyons. We are therefore furnished with the effective Lagrangian for tachyons. Recent works on noncommutative solitons~\cite{GhoshalSolitons} relied on the following deformation by star-products induced by a  term coupling the world-sheet to a $B$-field. This construction proceeded in a way that is inverse of the one we chose above in ordinary  (Archimedean) string theory. Tachyons of the $p$-adic string (or $p$-tachyons) were namely first considered as fields endowed with their anomalous dimensions, with no $B$-field turned on~\cite{Zabrodin}. The effective Lagrangian was subsequently deformed by coupling the boundary of the world-sheet to a magnetic field. We thus have got a theory of noncommutative $p$-tachyons with  built-in anomalous dimensions. Noncommutative tachyons of ordinary string were on the other hand considered in a limit where the anomalous dimensions vanish, and the present work proposed that anomalous dimensions should be incorporated as deformations of the noncommutative field theory, without destroying its structure. The toy-model of $p$-adic strings is expected to lead to ordinary strings when $p$ is sent to 1. This limit was successfully checked in the commutative set-up by Gerasimov and Shatashvili in~\cite{GS}, where it was observed that the effective action from boundary string field theory and the $p\rightarrow 1$ limit of $p$-adic strings lead to the same equations of motion upon a change of variables. Further checks could be related to  an interpretation of $B$ as a genuine closed-string background, instead of the  magnetic-field picture.\\

\noindent{{{ \Large \bf Acknowledgements}}}\\ 

\noindent{I} thank Marcos Mari\~no, Ruben Minasian and Pierre Vanhove for
discussions. This work was partially supported by EC Excellence Grant MEXT-CT-2003-509661.


\begin{thebibliography}{condensation}







 \bibitem{FT1} E.S. Fradkin and A.A. Tseytlin, \emph{Effective Field Theory
     from Quantized Strings}, 
 Phys. Lett. {\bf B158} (1985) 316.


\bibitem{FT2} E.S. Fradkin and A.A. Tseytlin, \emph{Nonlinear Electrodynamics from Quantized Strings}, Phys. Lett. {\bf{B163}} (1985) 123.



\bibitem{corrections}  O.D. Andreev and A.A. Tseytlin, \emph{Partition Function Representation For The Open Superstring Effective
Action: Cancellation Of M{\"o}bius Infinities And Derivative Corrections To
Born--Infeld Lagrangian},
Nucl. Phys.  {\bf B311}, 205 (1988).




\bibitem  {Wyllard}  N. Wyllard, \emph{ Derivative Corrections to D-brane Actions with Constant Background Fields}, Nucl. Phys. {\bf B598} (2001) 247-275, {\ttfamily hep-th/0008125}.



\bibitem {SW} N. Seiberg and E. Witten, \emph{String Theory and Noncommutative Geometry}, JHEP {\bf 9909} (1999) 032, {\ttfamily hep-th/9908142}.




\bibitem {DMS} S.R. Das, S. Mukhi and N.V. Suryanarayana, \emph{ Derivative Corrections from {\mbox{Noncommutativity}}}, JHEP {\bf 0108} (2001) 039, {\ttfamily hep-th/0106024}.




\bibitem{DT}  S.R. Das and S.P. Trivedi, \emph{ Supergravity Couplings
    to Non-Commutative Branes, Open Wilson Lines and Generalised Star Products}, JHEP {\bf 0102} (2001) 046, {\ttfamily hep-th/0011131}.



\bibitem{Liu} H. Liu, \emph{$*$-Trek II: $*_n$ Operations, Open Wilson Lines and the Seiberg--Witten Map}, {\mbox{Nucl. Phys.}} {\bf B614} (2001) 305-329, {\ttfamily hep-th/0011125}.



\bibitem {Mukhi} S. Mukhi, \emph{ Star Products from Commutative String Theory}, Pramana {\bf 58} (2002) 21-26, {\ttfamily hep-th/0108072}.


\bibitem {mezigue} P. Grange, \emph{Derivative Corrections from Boundary State Computations}, Nucl. Phys. {\bf
  B649} (2003) 297-311, {\ttfamily hep-th/0207211}.




\bibitem {MSbeyond} S. Mukhi and N.V. Suryanarayana, \emph{Open-String
    Actions and Noncommutativity {\mbox{Beyond}} the Large-B Limit},
  JHEP {\bf 0211} (2002) 002, {\ttfamily hep-th/0208203}.



\bibitem {Gbeyond} P. Grange, \emph{Modified Star-Products Beyond the
    Large-$B$ Limit}, Phys. Lett. {\bf B586} (2004) 125-132, {\ttfamily hep-th/0304059}.

\bibitem{condensation} A. Sen and B. Zwiebach, \emph{Tachyon condensation in string field theory}, JHEP {\bf 0003} (2000) 002, {\ttfamily{hep-th/9912249}}.






\bibitem {Shatashvili1} S.L. Shatashvili, \emph{Comment on the
    Background Independent Open String Theory},
  Phys. Lette. {\bf B311} (1993)83-86, {\ttfamily hep-th/9303143}.

\bibitem {Shatashvili2} S.L. Shatashvili, \emph{On the Problems with
    Background Independence in String Theory}, Alg. Anal. {\bf 6}
  (1994) 215-226,  {\ttfamily hep-th/9311177}.

\bibitem {GS} A.A. Gerasimov,  S.L. Shatashvili, \emph{On Exact Tachyon Potential in Open String Field Theory},  JHEP {\bf 0010} (2000) 034, {\ttfamily hep-th/0009103}.



\bibitem {KMM1} D. Kutasov, M. Mari{\~n}o and G. Moore, \emph{Some Exact
    Results on Tachyon Condensation in String Field Theory}, {\ttfamily hep-th/0009148}.

\bibitem {KMM2} D. Kutasov, M. Mari{\~n}o and G. Moore, \emph{Remarks on Tachyon Condensation in Superstring Field Theory}, {\ttfamily hep-th/0010108}.




\bibitem{DMR} K. Dasgupta, S. Mukhi, G. Rajesh, \emph{Noncommutative Tachyons},  JHEP {\bf 0006} (2000) 022, {\ttfamily hep-th/0005006}.


\bibitem{WittenNT} E. Witten, \emph{Noncommutative Tachyons And String Field Theory}, {\ttfamily hep-th/000607}.



\bibitem{Cornalba} L. Cornalba, \emph{Tachyon Condensation in Large
    Magnetic Fields with Background Independent String Field Theory}, {\ttfamily hep-th/0010021}.   


\bibitem {Okuyama} K. Okuyama, \emph{Noncommutative Tachyon from
    Background Independent Open String Field Theory}, Phys. Lett. {\bf
    B499} (2001) 167-173, {\ttfamily hep-th/0010028}.



\bibitem{Witten} E. Witten, \emph{Some Computations in Background Independent Open-String Field Theory}, Nucl.Phys. {\bf{B407}} (1993) 637-666, {\ttfamily hep-th/9307038}.


\bibitem{CFGNO}  E. Coletti, V. Forini, G. Grignani, G. Nardelli,
  M. Orselli, \emph{Exact potential and scattering amplitudes from
    the tachyon non-linear $\beta$-function}, JHEP {\bf 0403} (2004) 030, {\ttfamily hep-th/0402167}. 


\bibitem{KS} I. Klebanov and L. Susskind, \emph{Renormalization Group
  and String Amplitudes}, Phys. Lett. {\bf B200} (1988) 446.



\bibitem{universality} A. Sen, \emph{Universality of the Tachyon Potential}, JHEP {\bf 9912} (1999) 027,  {\ttfamily{hep-th/9911116}}.



\bibitem{GMS} R. Gopakumar, S. Minwalla and A. Strominger,
  \emph{Noncommutative Solitons}, JHEP {\bf 0005} (2000) 020, {\ttfamily hep-th/0005031}. 


\bibitem{GhoshalSolitons} D. Ghoshal, \emph{Exact Noncommutative Solitons in p-adic Strings and BSFT}, {\ttfamily hep-th/0406259}.





\bibitem{padic} P. Grange, \emph{Deformation of p-adic String Amplitudes in a Magnetic Field}, {\ttfamily{hep-th/0409305}}.




\bibitem{towards}   D. Ghoshal and  T. Kawano, \emph{Towards p-Adic String in Constant B-Field}, {\ttfamily{hep-th/0409311}}.


\bibitem{Zabrodin} A.V. Zabrodin, \emph{Non-Archimedean Strings and
    Bruhat--Tits Trees}, {\hbox{Comm. Math. Phys. {\bf 123}}} (1989) 463-483.


\end{thebibliography}
\end{document}